\documentclass[aip,prb,twocolumn]{revtex4}

\newcommand{\erfc}{\mbox{erfc}}
\newcommand{\erf}{\mbox{erf}}
\newcommand{\wolfp}{\alpha}
\renewcommand{\vec}{\mathbf}

\usepackage{graphics}
\usepackage{graphicx}

\begin{document}

\title{Simplistic Coulomb forces in molecular dynamics: Comparing the
  Wolf and shifted-force approximations}

\author{Jesper S. Hansen}
\email{jschmidt@ruc.dk}
\affiliation{
  DNRF Centre ``Glass and Time'', IMFUFA, 
  Department of Science, Systems and Models, Roskilde University, Postbox 260, 
  DK-4000 Roskilde, Denmark
}

\author{Thomas B. Schr{\o}der}
\email{tbs@ruc.dk}
\affiliation{
  DNRF Centre ``Glass and Time'', IMFUFA, 
  Department of Science, Systems and Models, Roskilde University, Postbox 260, 
  DK-4000 Roskilde, Denmark
}

\author{Jeppe C. Dyre}
\email{dyre@ruc.dk}
\affiliation{
  DNRF Centre ``Glass and Time'', IMFUFA, 
  Department of Science, Systems and Models, Roskilde University, Postbox 260, 
  DK-4000 Roskilde, Denmark
}

\begin{abstract}
This paper compares the Wolf method to the shifted forces (SF) method 
for efficient computer simulation of isotropic systems interacting via
Coulomb forces, taking results from the Ewald summation method as
representing the true behavior. We find that for the
Hansen-McDonald molten salt model the SF approximation overall
reproduces the structural and dynamical properties as accurately as
does the Wolf method. It is 
shown that the optimal Wolf damping parameter depends on the property
in focus, and that neither the potential energy nor the radial
distribution function are useful measures for the convergence of the
Wolf method to the Ewald summation method. The SF approximation is also
tested for the SPC/Fw model of liquid water at room temperature, showing
good agreement with both the Wolf and the particle mesh Ewald methods;
this confirms previous findings [Fennell \& Gezelter,
J. Chem. Phys. {\bf 124}, 234104 (2006)]. Beside its conceptual
simplicity the SF approximation implies a speed-up of a factor 2 to 3 
compared to the Wolf method (which is in turn much faster than the
Ewald method).  
\end{abstract}

\maketitle

\section{Introduction}
In molecular dynamics simulations the force evaluation consumes by far
the most computational resources. For relatively short-ranged
interactions like van der Waals interaction~\cite{mcquarrie_1976} it
is common to introduce a cutoff radius $r_c$ such that if the distance
between a particle pair exceeds $r_c$, the particles do not
interact~\cite{allen_1989}. This truncation allows for different
optimization methods like inclusion of cell and neighbor
lists~\cite{allen_1989, frenkel_1996, rapaport_1995}, 
which increase computational performance
considerably. Traditionally, the pair potential is simply truncated
and shifted such that it is zero at $r_c$
\cite{allen_1989,frenkel_1996, rapaport_1995}. This does not affect
the force acting between particles at distances below $r_c$, and if
$r_c$ is sufficiently large, the fluid properties are virtually
unaffected by this approximation. In fact, it has been
shown~\cite{weeks_1971,mcquarrie_1976} that keeping merely the
short-ranged, purely repulsive part of the van der Waals interaction
can account for the fluid structure even near the critical
point where correlations are long ranged. The truncated and shifted
potential approximation ensures continuity of the potential energy,
but introduces a discontinuity in the force at $r_c$, leading to
energy drift for long simulation times~\cite{toxvaerd_2011}. 
To overcome this one can instead apply a truncated and shifted force (SF)
approximation~\cite{allen_1989}, which has superior numerical
stability~\cite{toxvaerd_2011}. Beside the numerical stability, it was
recently shown by Toxvaerd and Dyre~\cite{toxvaerd_2011} that 
for highly dense fluids the SF method allows for 
very small cut-off radius of $r_c=1.5 \sigma$ (where $\sigma$ is
the atomic diameter) and corresponds approximately to the first local
minimum in the radial distribution function. Applying such low cutoff
to the truncated and shifted potential will lead to wrong physics and 
large energy drift~\cite{toxvard_2011}. The SF method therefore
decreases  the number of interactions significantly and thus the
computational time. The potential corresponding to the SF interaction 
does however not match the original potential for $r<r_c$ from which the
SF interaction was derived. Therefore, the thermodynamical
properties cannot be compared directly, but can be derived from
perturbation theory~\cite{nicolas_1979, allen_1989}. 

For long-ranged interactions, like the Coulomb interaction, one cannot
simply introduce a standard cut and shifted potential. For example,
simply truncating and shifting the Coulomb potential produces spurious
fluid structure and wrong dynamics~\cite{feller_1996}. Numerous
attempts have been made to overcome this problem. For example it has
been suggested to use smoothing functions, but this leads in general
to poor results, see Refs.~\onlinecite{brooks_1985,
  zahn_2002}. Wolf \emph{et al.}~\cite{wolf_1999} cleverly showed that
using a simple truncated and shifted Coulomb potential corresponds in
practice to summing over interactions in a non-neutral sphere. To
compensate for this these authors introduced a neutralizing term into
the Coulomb potential; they further showed that faster convergence to
the true energy is achieved by applying a damping factor $\wolfp$. The
Wolf method is computationally much faster than the classical Ewald summation
technique and is today widely used within the scientific simulation
community. The choice of the damping factor, $\wolfp$, is, like the
Ewald damping parameter~\cite{ewald_1921,allen_1989}, somewhat
arbitrary, and the optimal value must be found by comparison with
either experimental data or results from, e.g., the Ewald
method~\cite{zahn_2002,demontis_2001}.  If the Wolf damping parameter
$\wolfp$ is zero, the Wolf method reduces to the SF
approximation~\cite{fennell_2006}, see also Denesyuk and
Weeks~\cite{denesyuk_2008} for a discussion. We note that an SF method
for the Coulomb interactions was used as a clever trick in the
biochemical simulation community~\cite{levitt_1995,beck_2005} before
the work by Wolf \emph{et al.}. 

In this paper we apply the Wolf method in molecular dynamics
simulations of a simple model of a molten salt and liquid water. In
order to find the optimal value of the Wolf damping parameter $\wolfp$
we compare the simulated thermodynamical, dynamical, and structural
properties with previously published results~\cite{hansen_1975} based
on the Ewald method. We show that the optimal value of $\wolfp$
depends on the property one wishes to calculate and the cutoff
distance used. This sets the stage for documenting the main conclusion of
this paper: for the systems studied here the SF approximation works as
well as the Wolf method, confirming similar findings of Fennell and
Gezelter~\cite{fennell_2006}. Besides being conceptual simpler than
the Wolf method, the SF method allows for more than a doubling of the
computational speed.

\section{The Wolf approximation to the Coulomb potential}
If $r$ is the distance between two particles, the force acting on one
particle from the other is $\vec{F}(r) = f(r)\vec{r}/r$, where $f$
is for simplicity denoted the ``force'' and is minus the derivative of
the corresponding potential function with respect to $r$. For the Wolf
method~\cite{wolf_1999} the force is given by    
\begin{eqnarray}
  \nonumber
  f_W(r;\wolfp, r_c) &=& z_iz_j\left[
    \frac{\erfc(\wolfp r)}{r^2} - 
    \frac{\erfc(\wolfp r_c)}{r_c^2} \right. \\
   & + &\left. \frac{2\wolfp}{\sqrt{\pi}}
    \left( \frac{\exp(-\wolfp^2 r^2)}{r} -
      \frac{\exp(-\wolfp^2 r_{c}^2)}{r_{c}} 
    \right) 
  \right] \, \,  
  \label{eq:wolf}
\end{eqnarray}
for $r<r_c$ and where $\erfc(x) = 1 - \erf(x)$ is the complementary
error function. Here $z_i$ and $z_j$ are the charges of the two
particles in question, $r_c$ is the cutoff (i.e., $f_W=0$ for $r
\geq r_c$), and $\wolfp$ 
is the Wolf damping parameter. In the paper by Wolf {\it et al.} it is
implicitly understood that $\wolfp r_c>1$ such that the cutoff
only takes effect beyond range of damping. The damping parameter was
introduced in order to ensure faster convergence to the limiting
Madelung energy~\cite{wolf_1999}. Unfortunately, there is no
theoretical prediction for the optimal value of $\wolfp$, which must
be found by comparison with other well-established methods like the
Ewald summation method~\cite{wolf_1999,
  zahn_2002,demontis_2001}. Wolf \emph{et al.}~\cite{wolf_1999} and
Demontis \emph{et al.}~\cite{demontis_2001} have shown via molecular
dynamics simulations that the Wolf method reproduces the results
obtained by the Ewald summation method for $r_c\geq 5 d_{ij}$, where
$d_{ij}$ is the distance between oppositely charged particles in the
first coordinate shell. Demontis \emph{et al.}~\cite{demontis_2001}
also suggested that the optimal damping parameter is given by $\wolfp =
2/r_c$ for sufficiently large systems. 

From Eq. (\ref{eq:wolf}) it
follows that for $\wolfp \rightarrow \infty$ one has $f_W \rightarrow
0$, and that for $\wolfp\rightarrow 0$ the force reduces to   
\begin{eqnarray}
  f_{SF}(r; r_c) = z_iz_j\left(1/r^2 - 1/r_c^2 \right) \, \, \mbox{for}
  \, \, r<r_c . 
  \label{eq:sf}
\end{eqnarray}
This is the truncated and shifted force (SF) cutoff \cite{allen_1989,toxvaerd_2011}.  

\begin{figure}
  \begin{center}
    \includegraphics[scale=0.31]{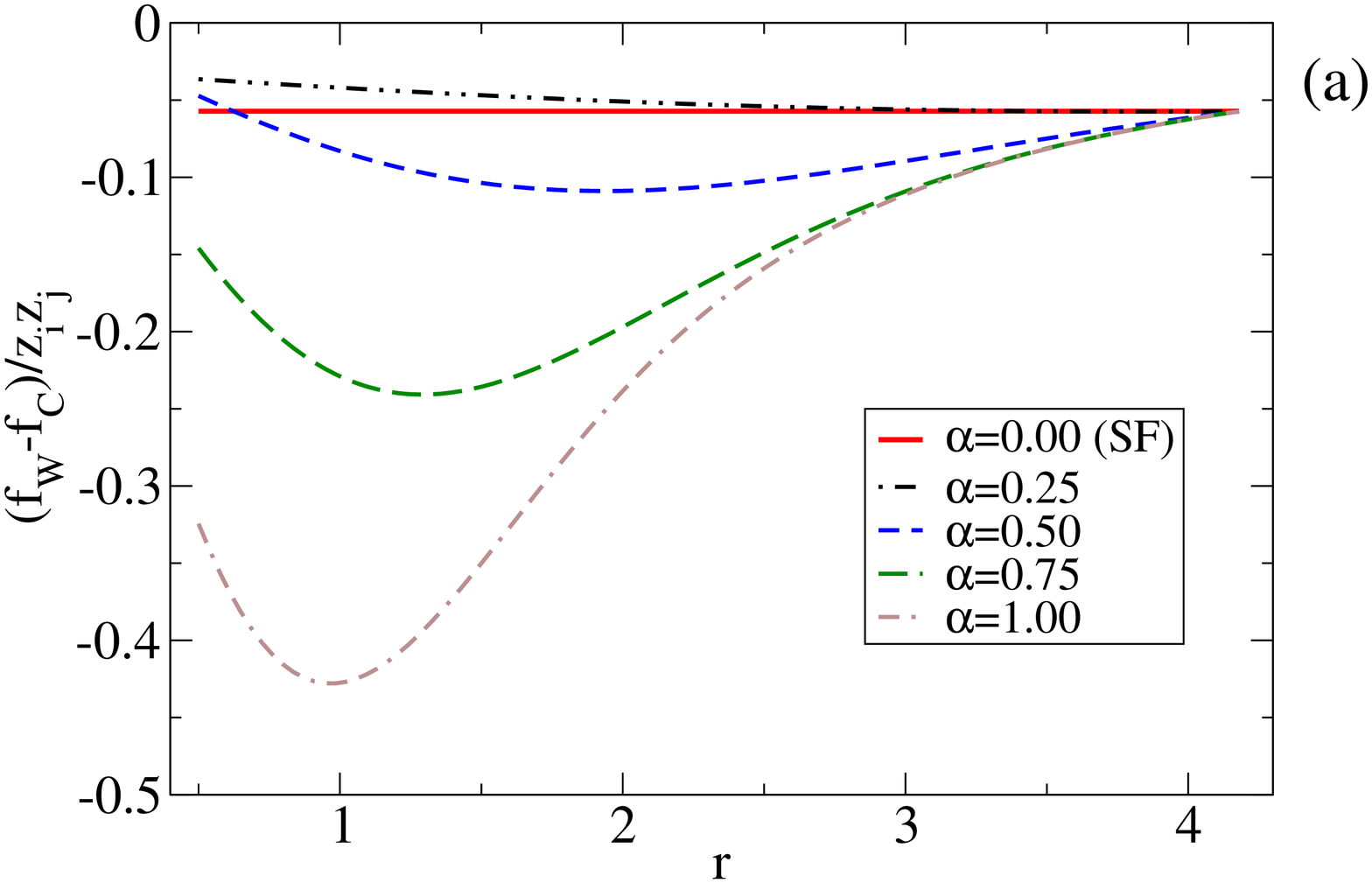}
    \\
    \includegraphics[scale=0.31]{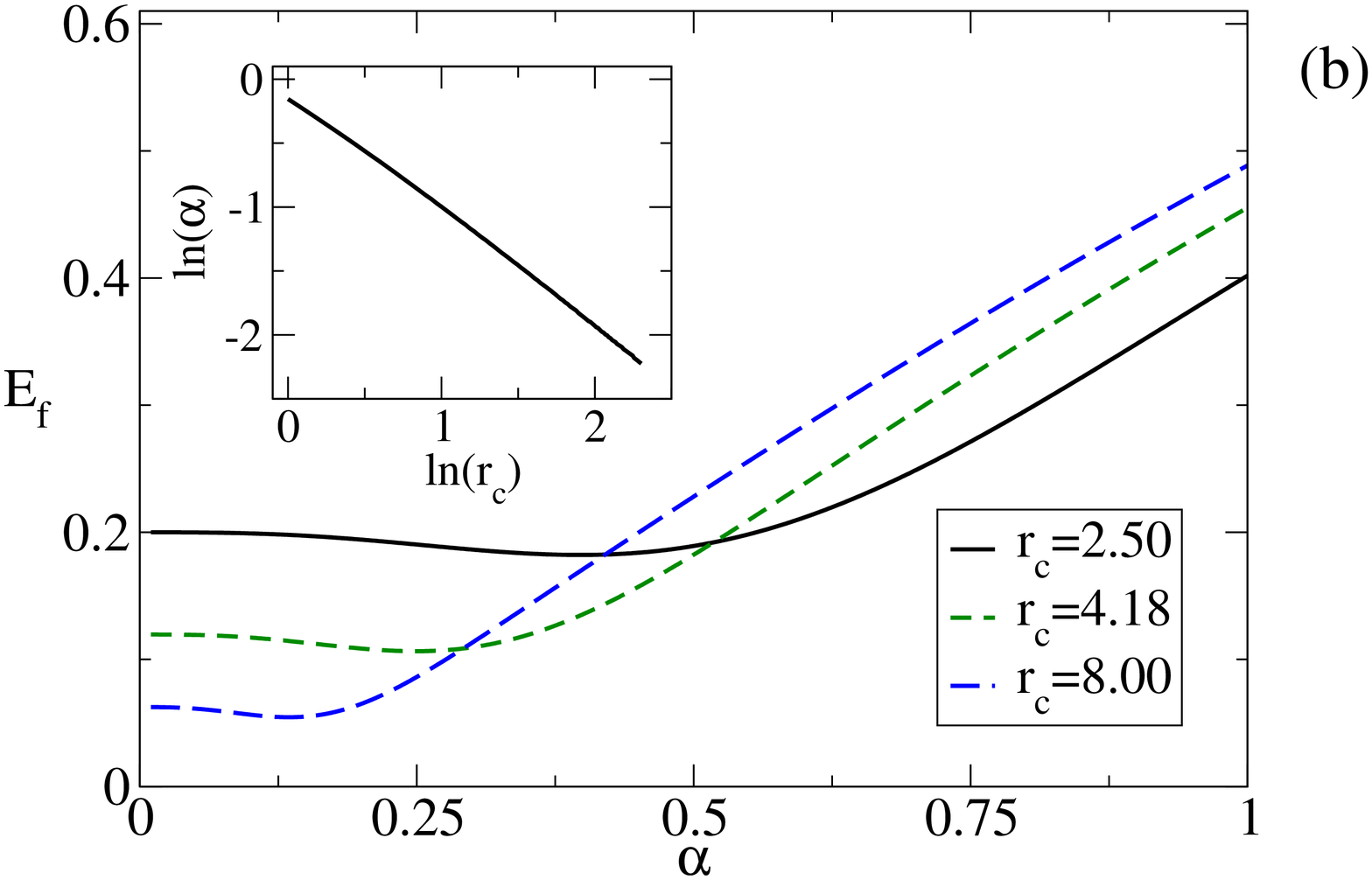}
  \end{center}
  \caption{\label{fig:0}
    [Color online] 
    (a): Difference between the Wolf force,
    $f_{\text{W}}$, and the Coulomb force, $f_C$, for $\wolfp=0.0,
    0.25, 0.5, 0.75$ and 1.0. In all graphs the cutoff is given by $r_c=4.18$. 
    (b): The measure of the difference between the true Coulomb force
    and the Wolf force, $E_f$, defined in Eq. (\ref{eq:minerr}) plotted
    as a function of $\wolfp$ for three different cutoffs. The inset
    shows the optimal value of $\wolfp$ plotted as a function of $r_c$.  
  }
\end{figure}

In Fig. \ref{fig:0}(a) we plot the difference between the Wolf
force, $f_W$, and the corresponding Coulomb force, $f_C
=z_iz_j/r^2$ , for different damping parameters.    
Clearly the damping parameter has a non-trivial effect on the
force. For $\wolfp=0$ the difference is small compared to large
values of $\wolfp$, suggesting that the SF cutoff, Eq.~(\ref{eq:sf}),
gives a good approximation to the Coulomb interaction. From
Fig.~\ref{fig:0}(a) it is seen that an optimal value of $\wolfp$
exists that minimizes the difference. One  way to identify this
optimal value is by minimizing the function
\begin{eqnarray}
\label{eq:minerr}
E_f(\wolfp, r_c) = 1 - \frac{\displaystyle \int_0^{r_c}f_W(r;\wolfp,
r_c) \, \mbox{d}r}{\displaystyle \int_0^{r_c} f_C(r) \,
\mbox{d}r} ,
\end{eqnarray}
which measures the total relative difference between $f_W(r)$ and
$f_C(r)$ such that $E_f \geq 0$ (since $f_W \leq f_C$ for all $r$). 
In Fig. \ref{fig:0}(b) $E_f$ is plotted for three different
cutoff distances. The optimal Wolf damping parameter converges to zero
as $r_c$ increases, which reflects the simple fact that $f_W \rightarrow f_C$
for $r_c \rightarrow \infty$ and $\wolfp \rightarrow 0$. More
interestingly, the quantity $E_f$ does exhibit very little 
difference between the optimal value of $\wolfp$ and $\wolfp=0$. The
inset in Fig. \ref{fig:0}(b) shows that the optimal Wolf parameter
determined by the minimum of Eq.~(\ref{eq:minerr}) is given roughly by
$\wolfp \approx 3/(4 r_c)$. This simple analysis is consistent with
the $r_c$ dependence suggested by Demontis \emph{et
  al.}~\cite{demontis_2001} based of molecular dynamics simulations
(but they predict a smaller estimate of $\wolfp$ by a factor of 3/8).       

The conclusion from Fig. \ref{fig:0} is that setting $\wolfp=0$, i.e.,
adopting the SF approximation, gives results that are close to those
obtained by carefully optimizing $\wolfp$. This and the recent work by
Toxvaerd and Dyre~\cite{toxvaerd_2001} motivate the below
reported molecular dynamics simulations, which compare the Wolf method
to the SF cutoff for other quantities and realistic systems. As
``truth'' we take the well-established, but computationally expensive,
Ewald summation method.

\section{Results for the Hansen and McDonald molten salt model}

A series of molecular dynamics simulations was performed of a model molten salt
proposed by Hansen and McDonald~\cite{hansen_1975}. 
Briefly, in this two-component model the ions are simple
spherical particles that interact via a Coulomb potential and 
a van der Waals type potential given by the inverse power law $\phi(r) =
\frac{\epsilon^2}{n\sigma}\left(\frac{\sigma}{r}\right)^n$, where
$n=9$, $\epsilon$ defines the energy scale and $\sigma$ is the usual
Lennard-Jones length scale parameter~\cite{mcquarrie_1976}.
We refer the reader to the reference for the full details. In the
simulations we applied the Wolf method and varied the cutoff between
2.5 and 8.0 $\sigma$. The simulation box used was twice the
size of the cutoff whenever $r_c>4.18 \, \sigma$; for smaller cutoffs
the box length was fixed to 8.36 $\sigma$. The number density for all
systems were $\rho=0.368\sigma^{-3}$, thus, the number of ions varied
from 216 to 1508. The results presented below were found to be
independent of system size. The temperature $T$ is controlled using a
Nos\'{e}-Hoover thermostat~\cite{nose_1984,hoover_1985} with 
$T= 0.0177 \epsilon/k_B$. The results are compared
to previously published data where the Ewald summation method was
used~\cite{hansen_1975}, which represent the ``true'' Coulomb
interaction.     
 
\begin{figure}
  \begin{center}
    \includegraphics[scale=0.31]{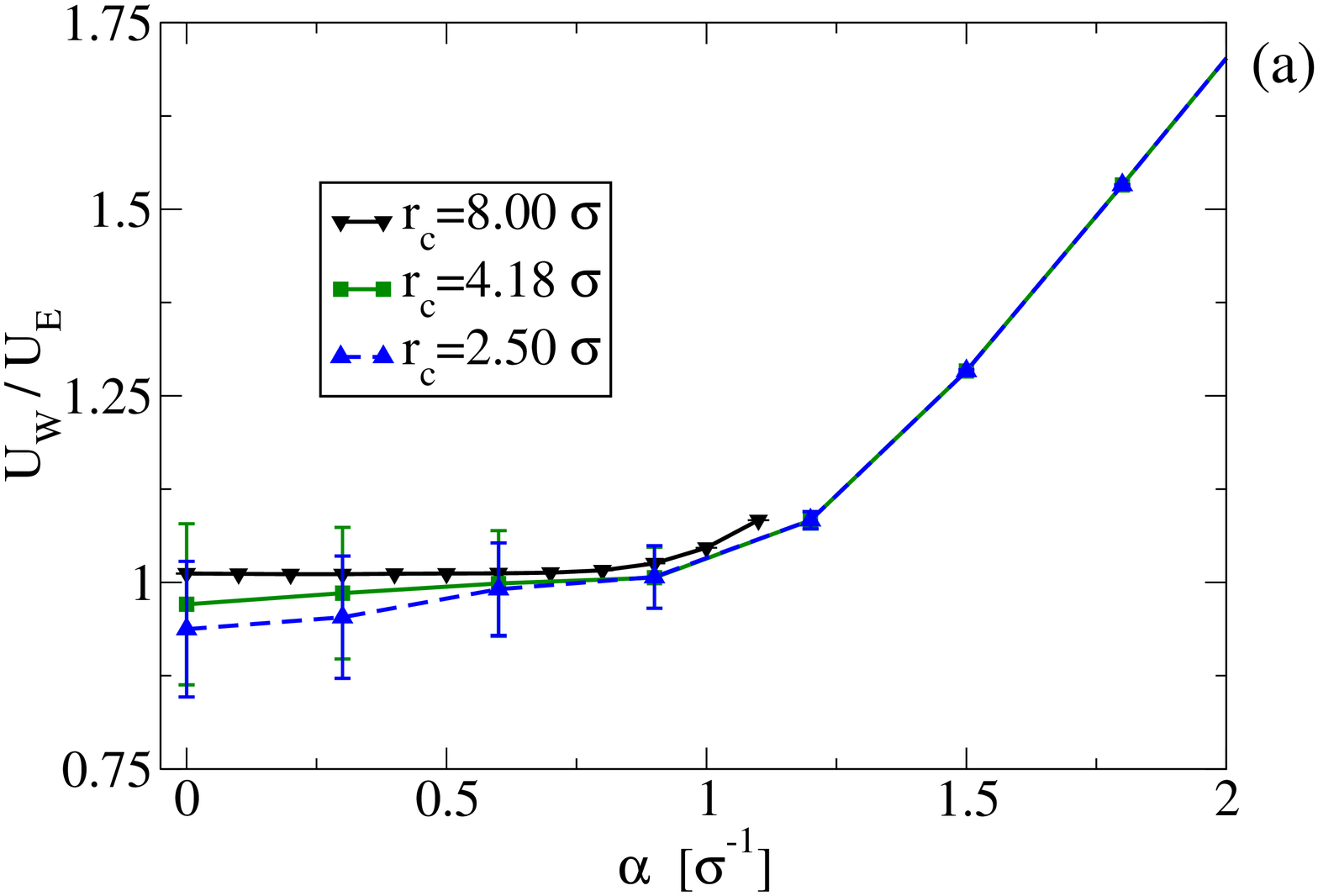}
    \\
    \includegraphics[scale=0.31]{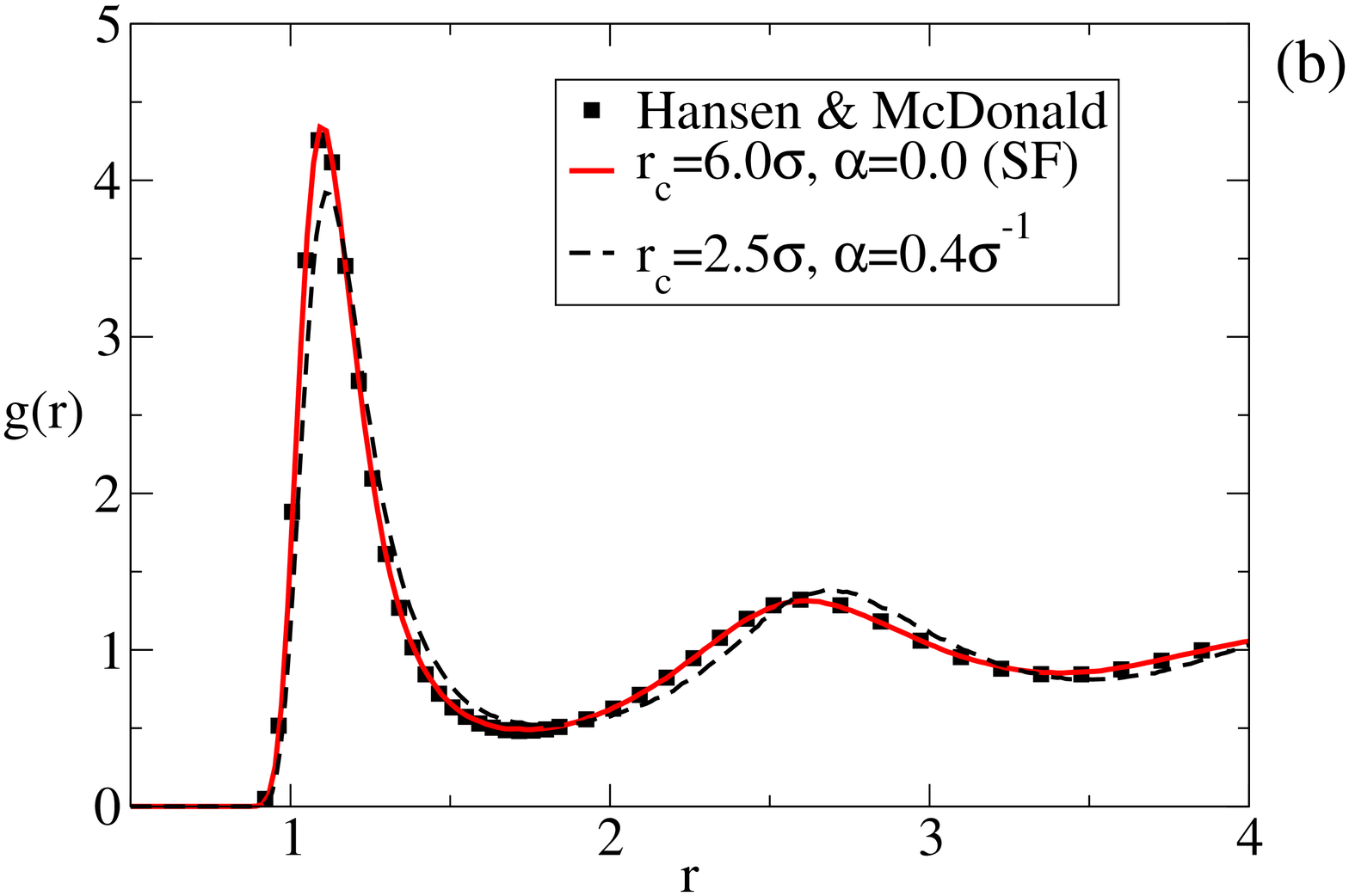}
  \end{center}
  \caption{\label{fig:1}
    [Color online] 
    (a): Comparison of the potential energy of the Hansen-McDonald
    molten salt model for varying  damping parameter. Error bars
    represent the standard error of ten 
    independent runs. $U_E$ is found in Ref.~\onlinecite{hansen_1975}.
    (b): Radial distribution functions for unlike charged particles 
    (lines) for $\wolfp = 0$
    (the SF approximation) and $r_c=6 \sigma$ and for $\wolfp=0.4
    \sigma^{-1}$ and $r_c=2.5 \sigma$. The filled black squares are
    data points taken from Ref.~\onlinecite{hansen_1975}. 
  }
\end{figure}

First, in Fig.~\ref{fig:1} (a) we compare the total potential energy
obtained from the Wolf method $U_W$ for three different cutoff radii
and varying damping parameters with the potential energy $U_E$ from
the Ewald summation method. We note that $U_W$ is obtained directly
from the Wolf potential function\cite{wolf_1999,zahn_2002}
corresponding to the force given in Eq. (\ref{eq:wolf}).  
It is observed that $U_W$ is within the statistical uncertainty equal
to $U_E$ for sufficiently small damping parameters, even for quite
small cutoffs. This could lead to the conclusion that the Wolf
method accounts correctly for electrostatic interactions for small
cutoff distances. However, if one plots the radial distribution function $g$,
Fig.~\ref{fig:1}(b), we see that for $r_c = 2.5 \sigma$ the
structure differs significantly from  the result obtained using the
Ewald summation method. This is true for all values of the damping
parameter. From Fig.~\ref{fig:1}(b) we also notice that the SF
approximation captures the structural properties correctly for $r_c =
6 \sigma$, which is the smallest cutoff distance meeting the Wolf 
\emph{et al.}~\cite{wolf_1999} and Demontis \emph{et
  al.}~\cite{demontis_2001} criterion $r_c \geq 5d_{ij}$.   

\begin{figure}
  \begin{center}
    \includegraphics[scale=0.31]{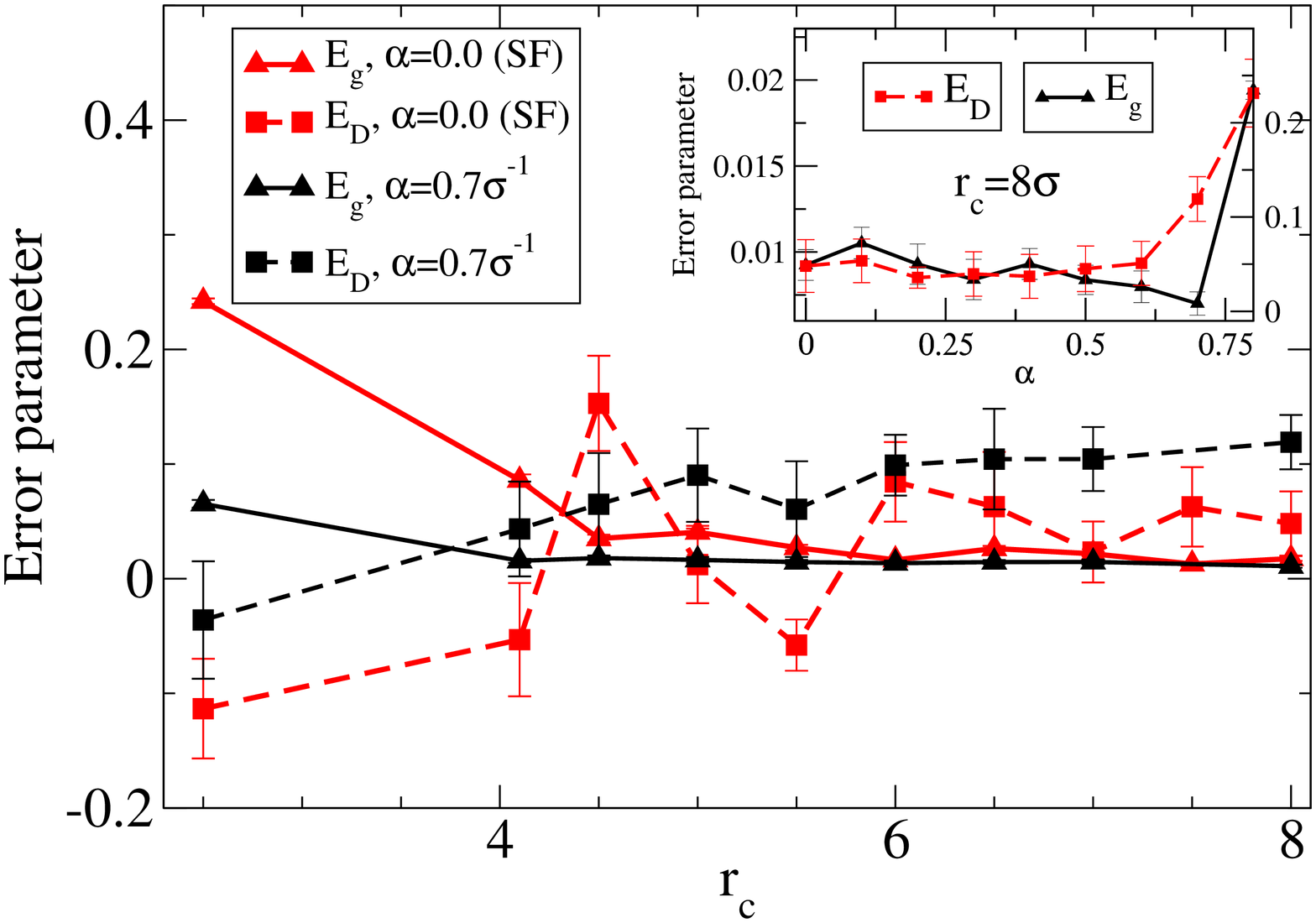}
  \end{center}
  \caption{\label{fig:2}
    [Color online] 
    Error parameters as a function of cutoff for
    different damping parameter for the Hansen and McDonald molten
    salt system. The inset shows the error parameters
    for $r_c=8 \sigma$ as functions of $\wolfp$.}
\end{figure}

We study the radial distribution function dependence of $r_c$ and
$\wolfp$ by defining the error parameter $E_g$ via 
\begin{eqnarray}
E_g =  \frac{\displaystyle \int_0^{r_c}
  |g_W(r)-g_E(r)| \, \mbox{d}r}{\displaystyle \int_0^{r_c} g_E(r)
  \, \mbox{d}r}
\ , \label{eq:p1}
\end{eqnarray}
where $g_W$ is the radial distribution function for unlike charged
particles of the Wolf method and $g_E$ the radial distribution
function produced by the Ewald summation method. Similarly, the
following error parameter $E_D$ quantifies the difference in diffusion
constant 
\begin{eqnarray}
E_D = \frac{D_W}{D_E} - 1 \ , \label{eq:p2}
\end{eqnarray}
where $D_W$ and $D_E$ are the diffusion constants obtained from the
Wolf and Ewald methods, respectively. Note that $E_g\ge 0$, whereas
$E_D$ can be negative. The ``correct'' radial distribution function,
$g_E$, and diffusion constant, $D_E$, were taken from
Ref.~\onlinecite{hansen_1975}. Figure \ref{fig:2} shows the two error
parameters for different cutoff radii and damping.  The damping
parameter $\wolfp = 0.7 \sigma^{-1}$ was chosen because $E_g$ exhibits
a minimum for this value for a large range of cutoffs. This is not the
case for $E_D$, however, which features a minimum for lower
values of the damping parameter, depending on the cutoff (as expected
from Fig.~\ref{fig:0} (b)). This inconsistency is illustrated in the
inset in which the error parameters are shown for $r_c=8 \sigma$ as
functions of $\wolfp$. Obviously, any $\wolfp<0.6 \sigma^{-1}$ may be
chosen to minimize $E_D$, whereas $E_g$ features a minimum for
$\wolfp=0.7 \sigma^{-1}$. We note that $r_c=8 \sigma > 5d_{ij}$
and the cutoff radius fulfills the criterion defined by Wolf \emph{et al.} and
Demontis \emph{et al.}.   

From Fig. \ref{fig:2} it is seen that $E_g$ is relatively large
for small cutoffs (as expected), but that it for non-zero damping
parameters quickly decreases and reaches almost zero for $r_c>4.0
\sigma^{-1}$. For the SF approximation one needs $r_c>6.0\sigma$ in
order to obtain the same accuracy in the radial distribution
function. For large cutoffs the SF approximation results in better
diffusion constants than the Wolf method with
$\wolfp=0.7\sigma^{-1}$. We could, of course, have optimized $\wolfp$
with respect to the diffusion constant (giving $\wolfp \cong
 0.3 \sigma^{-1}$ for a large range of cutoffs).  
This, however, would decrease the agreement for the
radial distribution function. This fact is highlighted in Table I, where
the error parameters are listed for values of $\wolfp$ optimized,
respectively, with respect to the diffusion constant and the radial
distribution function ($r_c=8.0 \sigma$). For comparison we also give 
the error parameters for the SF approximation.  
\begin{table}
  \begin{center}
    \begin{tabular}{lcccc} \hline \hline 
      $\wolfp$ $[\sigma^{-1}$] & $\mbox{}$ & $E_D$           & $\mbox{}$ & $E_g$             \\ \hline 
      0.0  (SF)               &$\mbox{}$  & 0.04 $\pm$ 0.02 &$\mbox{}$  & 0.017 $\pm$ 0.002 \\
      0.3                     & $\mbox{}$ & 0.03 $\pm$ 0.01 &$\mbox{}$  & 0.019 $\pm$ 0.002 \\
      0.7                     & $\mbox{}$ & 0.12 $\pm$ 0.02 & $\mbox{}$ & 0.010 $\pm$ 0.001 \\
      \hline      \hline
    \end{tabular}
    \caption{\label{table:1}
      Error parameters, $E_D$ and $E_g$, for
      different values of the damping parameter.
      $\wolfp=0.3\sigma^{-1}$ and $\wolfp=0.7\sigma^{-1}$ correspond
      to the optimized values with respect to diffusion and
      radial distribution function, respectively. $\wolfp=0.0\sigma^{-1}$
      corresponds to the SF approximation.}
  \end{center}
\end{table}
Within the statistical uncertainty there is no difference
between the Wolf method using $\wolfp = 0.3 \sigma^{-1}$ and the SF
approximation. 

\begin{figure}
  \begin{center}
    \includegraphics[scale=0.31]{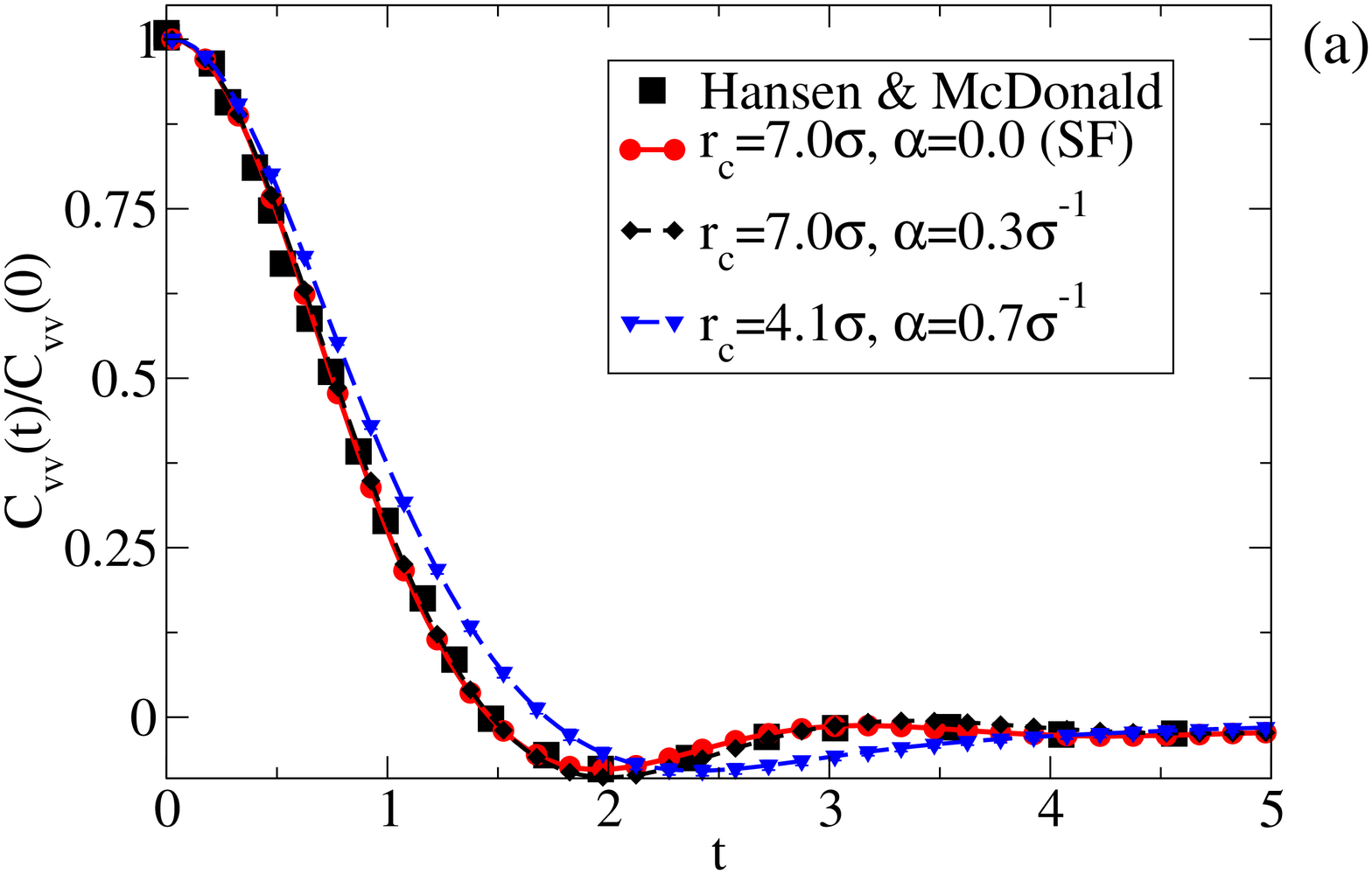}
    \\
    \includegraphics[scale=0.31]{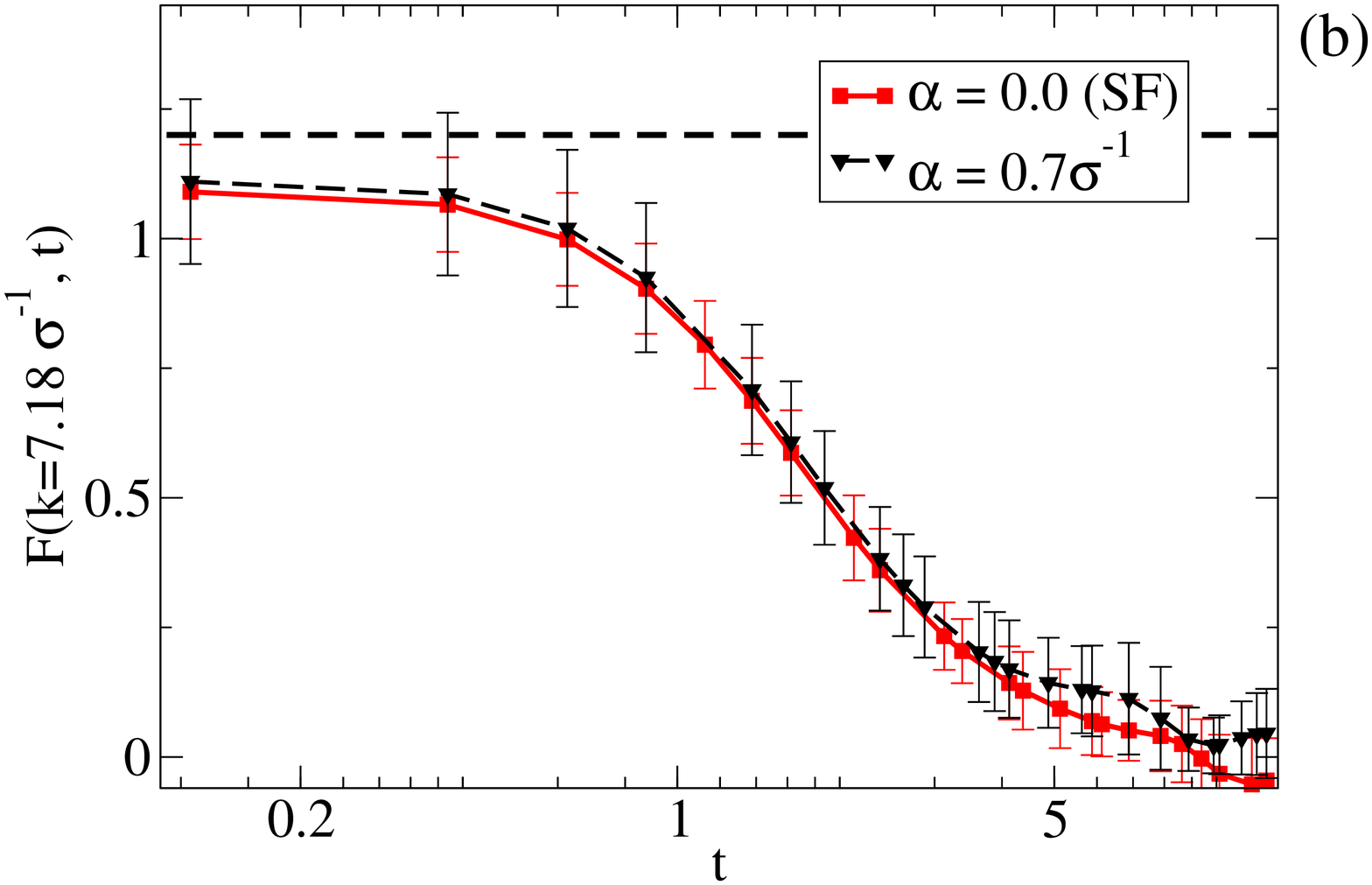}
  \end{center}
  \caption{\label{fig:3}
    [Color online]
    (a): Normalized velocity autocorrelation function for the Wolf and
    the SF methods. Only the short time data are shown. The error bars
    are comparable to the size of symbols. 
    (b): Coherent intermediate scattering function for wave-length $k=7.18
    \sigma^{-1}$ and $r_c=7.0\sigma$. The horizontal line is the
    interpolated value of the static structure factor $S(k)=F(k,0)$
    taken from Ref.~\onlinecite{hansen_1975}. The time $t$ is given in
    standard reduced molecular dynamics units.
  }
\end{figure}

Up to this point we have only discussed the structural and diffusive
properties in the long time limit. To compare the short-time dynamics
of the two methods we plot the velocity autocorrelation function $C_{vv}(t)$ and the
intermediate scattering function in Fig. \ref{fig:3}. 
From Figs. \ref{fig:1} and \ref{fig:2} it was concluded that for small
$r_c$ ($r_c \approx 4.0 \sigma$) and large $\wolfp$ ($\wolfp \approx
0.7$)  both the potential energy and the
radial distribution function are in excellent agreement with the Ewald
summation method, but in Fig.~\ref{fig:3} we clearly observe 
the short-time dynamics is not correct for this set of parameter
values. This shows that the cutoff must be sufficiently large for the
Wolf method to correctly account for all the fluid properties -- but
at such large cutoff the SF approximation may be applied instead since
it results in the same accuracy.

\section{Results for a water model}

\begin{figure}
  \begin{center}
    \includegraphics[scale=0.31]{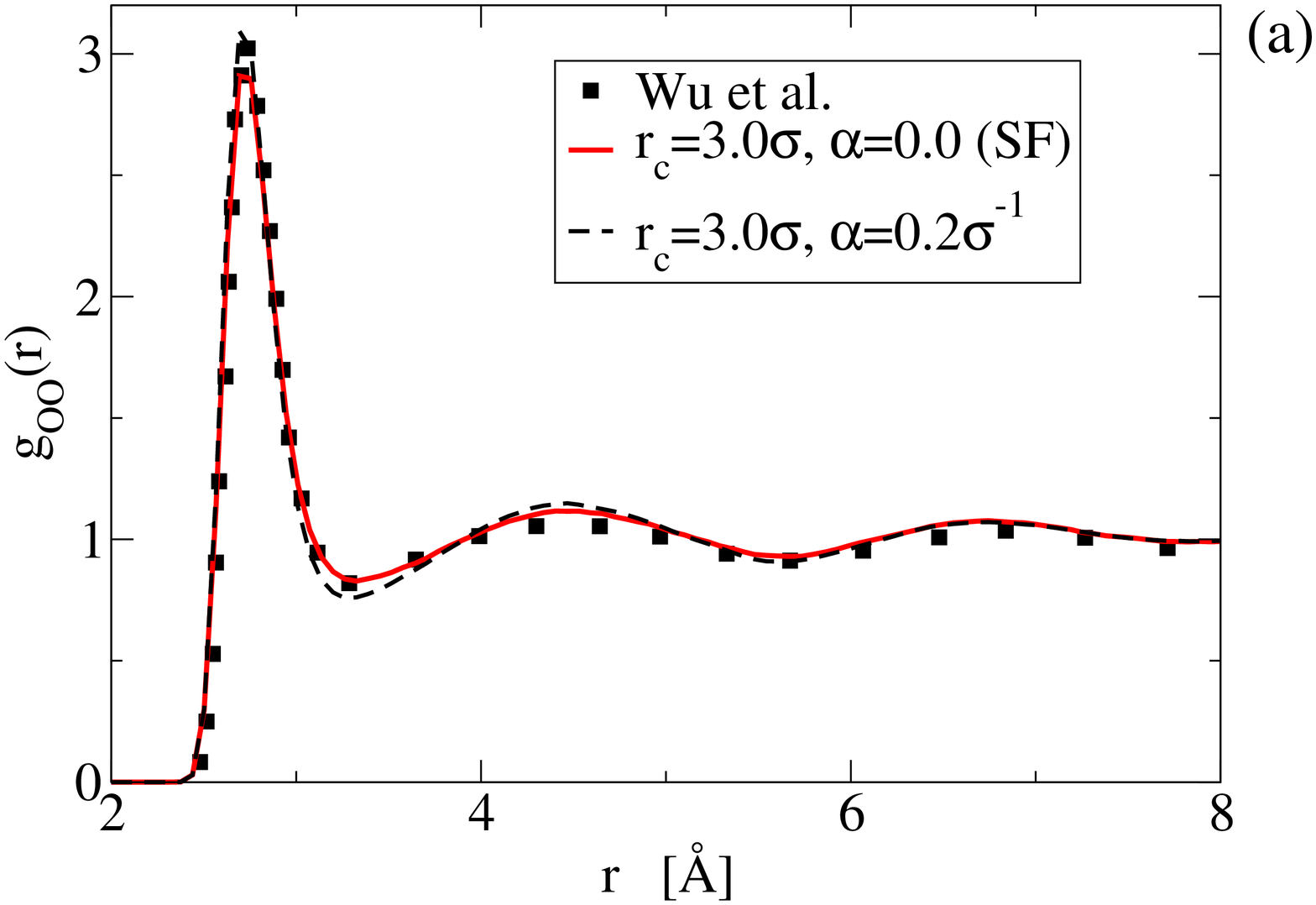}
    \\
    \includegraphics[scale=0.31]{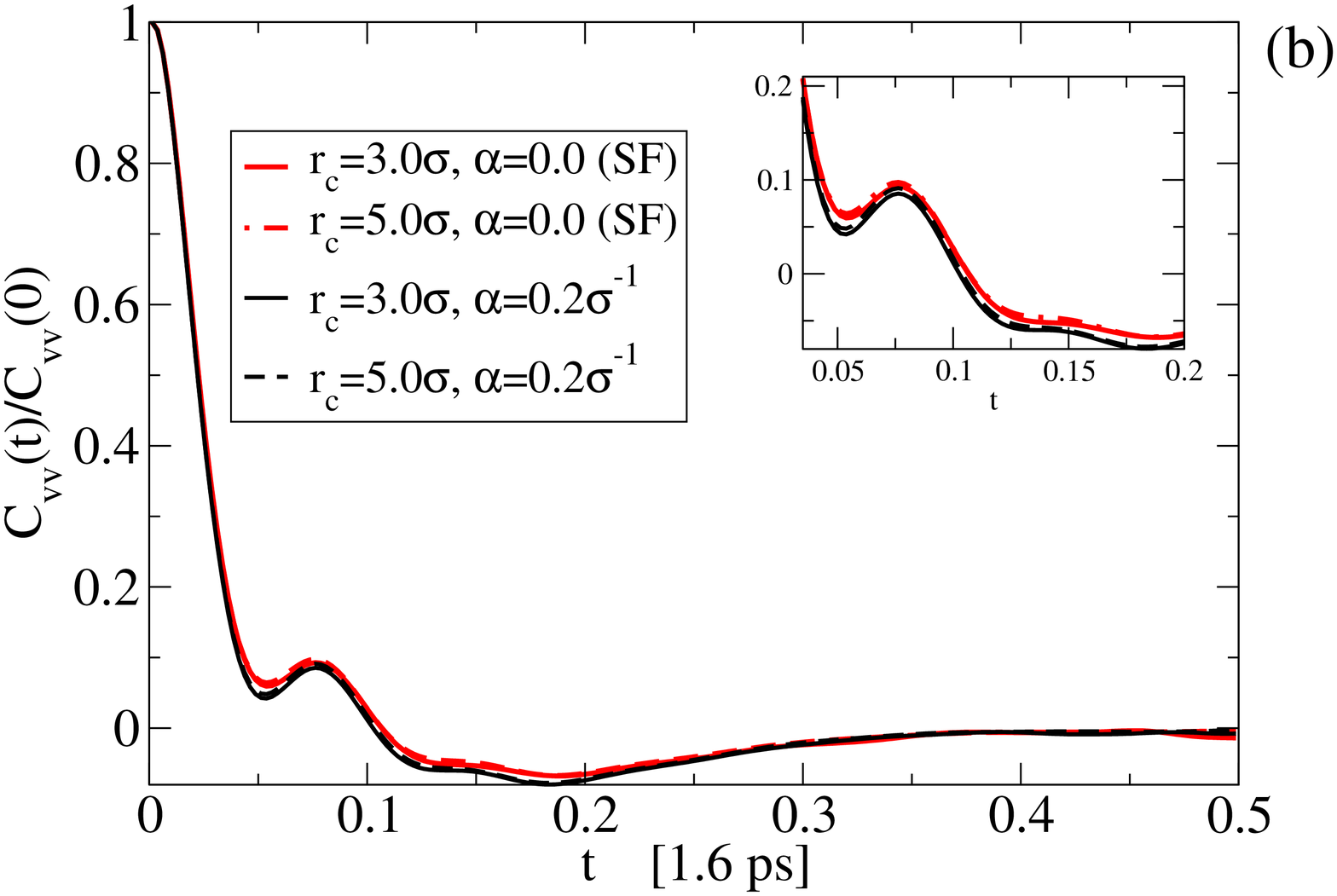}
  \end{center}
  \caption{\label{fig:4}
    [Color online]
    (a): Oxygen-oxygen radial distribution function for the SPC/Fw
    water model using SF and Wolf methods. The squares represent 
    data taken from  Ref.~\onlinecite{wu_2006}.
    (b): Normalized center-of-mass velocity autocorrelation
    functions. The inset shows a zoom of the time interval 0.065 to
    0.32 ps. In both (a) and (b) $\sigma = 3.16$ \AA.
  }
\end{figure}

We also tested the SF approximation for liquid water at the
state point $(T, \rho)$ = (300 K, 998 kg m$^{-3}$) using the
flexible single point charge (SPC/Fw) water model~\cite{wu_2006}. In
this model the chemical bond and the bending angle are allowed to
vibrate around their zero-force values. The model is easy to implement
and has been shown to predict many bulk properties better than for
example the SPC, SPC/E and TIP3P models~\cite{wu_2006, raabe_2007}. In
Fig. \ref{fig:4}(a) we plot the oxygen-oxygen radial distribution
function $g_{oo}$ for the Wolf and SF methods. For comparison, data
from Ref. \onlinecite{wu_2006} are shown (filled squares), where the
Coulomb interactions were evaluated using the particle-mesh Ewald
(PME) method~\cite{darden_1993}.  
The radial distribution
function is reproduced reasonably well by both methods. The SF
approximation captures the liquid structure at least as well as the
Wolf method, except at the first peak which is slightly underestimated. The
radial distribution functions for both the SF and the
Wolf methods are independent of the cutoff for radii larger than 9
{\AA}, the value used by Zahn \emph{et al.}~\cite{zahn_2002}; this
corresponds to $r_c\approx 5d_{ij}$ since the oxygen-hydrogen
distance is around 1.8 \AA. In Fig.~\ref{fig:4}(b) the center-of-mass
velocity autocorrelation function is plotted for two different cutoffs
for both methods. This dynamic property is largely  independent of
method and cutoff, as is the case for the liquid structure. The same
conclusion was reached by Fennell and
Gezelter~\cite{fennell_2006}. For the SF approximation we 
obtain a diffusion constant of 2.4 $\times 10^{-9}$ m$^2$ s$^{-1}$ and
a shear viscosity of 0.78 $\times 10^{-3}$ Pa s. This can be compared
with the experimental values 2.3 $\times 10^{-9}$ m$^2$ s$^{-1}$ and
0.85 $\times 10^{-3}$ Pa s. It is also worth mentioning
that Zahn \emph{et  al.}~\cite{zahn_2002} used $\wolfp=0.06
\sigma^{-1}$ in their simulations of (rigid) SPC/E water, but found
that the potential energy was in better agreement with the Ewald
method for even lower damping parameters. For $\wolfp=0.06
\sigma^{-1}$ we observe no difference to the SF approximation.

\section{Concluding remarks}
In conclusion, for simple molten salts and liquid water the SF
approximation reproduces various properties as well as the Wolf
method. The Wolf method has one more parameter than SF, and
consequently this method may be optimized to give slightly better
agreement with 
the Ewald summation method. Such an optimization, however, must be
carried separately out for each property in focus and for each
different system. Beside its simplicity (and thus easy-to-code
feature), we found that the SF approximation leads to a simulation
speed-up of 2-3 compared to the Wolf method. Of course, the actual speed-up
depends on the specific problem and the use of optimization
techniques, but the calculation of the four terms in
Eq. (\ref{eq:wolf}) involves complicated mathematical functions and is
deemed to consume considerably more computational resources than the
simple SF approximation. We wish to stress here that the paper of Wolf
\emph{et al.} was the first to correctly analyze why
the SF approximation for Coulomb forces is superior to the standard
truncated and shifted potential interaction model.  

Fennell and Gezelter~\cite{fennell_2006} carefully analyzed an
impressive number of different systems including simple crystals
showing a good agreement between the SF method and the Ewald
technique. In their conclusion the authors suggested that the SF
approximation can also be used for confined geometries, thereby
overcoming the enforced periodicity in the unmodified Ewald method. We
agree that the Ewald method can be problematic (even for periodic
systems~\cite{karlstrom_2008}), but, the SF approach (as well as
the Wolf method) is an approximation that suppresses the intrinsic
long-ranged nature of the Coulomb interactions leading to an
artificially molecular orientation~\cite{feller_1996,takahashi_2011}
in confinements. For confined fluids alternative methods have recently
been adviced, see Refs.~\onlinecite{rodgers_2008,wu_2005,denesyuk_2008}.  

\section{Acknowledgements}
JSH wishes to acknowledge Lunbeckfonden for supporting this work as
part of grant no. R49-A5634. The centre for viscous liquid dynamics
``Glass and Time'' is sponsored by the Danish National Research
Foundation (DNRF).


\begin{thebibliography}{27}
\expandafter\ifx\csname natexlab\endcsname\relax\def\natexlab#1{#1}\fi
\expandafter\ifx\csname bibnamefont\endcsname\relax
  \def\bibnamefont#1{#1}\fi
\expandafter\ifx\csname bibfnamefont\endcsname\relax
  \def\bibfnamefont#1{#1}\fi
\expandafter\ifx\csname citenamefont\endcsname\relax
  \def\citenamefont#1{#1}\fi
\expandafter\ifx\csname url\endcsname\relax
  \def\url#1{\texttt{#1}}\fi
\expandafter\ifx\csname urlprefix\endcsname\relax\def\urlprefix{URL }\fi
\providecommand{\bibinfo}[2]{#2}
\providecommand{\eprint}[2][]{\url{#2}}

\bibitem[{\citenamefont{Mcquarrie}(1976)}]{mcquarrie_1976}
\bibinfo{author}{\bibfnamefont{D.~A.} \bibnamefont{Mcquarrie}},
  \emph{\bibinfo{title}{Statistical Mechanics}} (\bibinfo{publisher}{Harper and
  Row}, \bibinfo{address}{New York}, \bibinfo{year}{1976}).

\bibitem[{\citenamefont{Allen and Tildesley}(1989)}]{allen_1989}
\bibinfo{author}{\bibfnamefont{M.~P.} \bibnamefont{Allen}} \bibnamefont{and}
  \bibinfo{author}{\bibfnamefont{D.~J.} \bibnamefont{Tildesley}},
  \emph{\bibinfo{title}{Computer {S}imulation of {L}iquids}}
  (\bibinfo{publisher}{Clarendon Press}, \bibinfo{address}{New York},
  \bibinfo{year}{1989}).

\bibitem[{\citenamefont{Frenkel and Smit}(1996)}]{frenkel_1996}
\bibinfo{author}{\bibfnamefont{D.}~\bibnamefont{Frenkel}} \bibnamefont{and}
  \bibinfo{author}{\bibfnamefont{B.}~\bibnamefont{Smit}},
  \emph{\bibinfo{title}{Understanding Molecular Simulation}}
  (\bibinfo{publisher}{Academic Press}, \bibinfo{address}{London},
  \bibinfo{year}{1996}).

\bibitem[{\citenamefont{Rapaport}(1995)}]{rapaport_1995}
\bibinfo{author}{\bibfnamefont{D.}~\bibnamefont{Rapaport}},
  \emph{\bibinfo{title}{The Art of Molecular Dynamics Simulation}}
  (\bibinfo{publisher}{Cambridge University Press},
  \bibinfo{address}{Cambridge}, \bibinfo{year}{1995}).

\bibitem[{\citenamefont{Weeks et~al.}(1971)\citenamefont{Weeks, Chandler, and
  Andersen}}]{weeks_1971}
\bibinfo{author}{\bibfnamefont{J.~D.} \bibnamefont{Weeks}},
  \bibinfo{author}{\bibfnamefont{D.}~\bibnamefont{Chandler}}, \bibnamefont{and}
  \bibinfo{author}{\bibfnamefont{H.~C.} \bibnamefont{Andersen}},
  \bibinfo{journal}{J. Chem. Phys.} \textbf{\bibinfo{volume}{54}},
  \bibinfo{pages}{5237} (\bibinfo{year}{1971}).

\bibitem[{\citenamefont{Toxvaerd and Dyre}(2011)}]{toxvaerd_2011}
\bibinfo{author}{\bibfnamefont{S.}~\bibnamefont{Toxvaerd}} \bibnamefont{and}
  \bibinfo{author}{\bibfnamefont{J.~C.} \bibnamefont{Dyre}},
  \bibinfo{journal}{J. Chem. Phys} \textbf{\bibinfo{volume}{134}},
  \bibinfo{pages}{081102} (\bibinfo{year}{2011}).

\bibitem[{\citenamefont{Nicolas et~al.}(1979)\citenamefont{Nicolas, Gubbins,
  Streett, and Tildesley}}]{nicolas_1979}
\bibinfo{author}{\bibfnamefont{J.~J.} \bibnamefont{Nicolas}},
  \bibinfo{author}{\bibfnamefont{K.~E.} \bibnamefont{Gubbins}},
  \bibinfo{author}{\bibfnamefont{W.~B.} \bibnamefont{Streett}},
  \bibnamefont{and} \bibinfo{author}{\bibfnamefont{D.~J.}
  \bibnamefont{Tildesley}}, \bibinfo{journal}{Mol. Phys.}
  \textbf{\bibinfo{volume}{37}}, \bibinfo{pages}{1429} (\bibinfo{year}{1979}).

\bibitem[{\citenamefont{Feller et~al.}(1996)\citenamefont{Feller, Pastor,
  Rojnuckarin, Bogusz, and Brooks}}]{feller_1996}
\bibinfo{author}{\bibfnamefont{S.~E.} \bibnamefont{Feller}},
  \bibinfo{author}{\bibfnamefont{R.~W.} \bibnamefont{Pastor}},
  \bibinfo{author}{\bibfnamefont{A.}~\bibnamefont{Rojnuckarin}},
  \bibinfo{author}{\bibfnamefont{S.}~\bibnamefont{Bogusz}}, \bibnamefont{and}
  \bibinfo{author}{\bibfnamefont{B.~R.} \bibnamefont{Brooks}},
  \bibinfo{journal}{J. Phys. Chem.} \textbf{\bibinfo{volume}{100}},
  \bibinfo{pages}{17011} (\bibinfo{year}{1996}).

\bibitem[{\citenamefont{Brooks et~al.}(1985)\citenamefont{Brooks, Pettitt, and
  Karplus}}]{brooks_1985}
\bibinfo{author}{\bibfnamefont{C.~L.} \bibnamefont{Brooks}},
  \bibinfo{author}{\bibfnamefont{B.~M.} \bibnamefont{Pettitt}},
  \bibnamefont{and} \bibinfo{author}{\bibfnamefont{M.}~\bibnamefont{Karplus}},
  \bibinfo{journal}{J. Chem. Phys.} \textbf{\bibinfo{volume}{83}},
  \bibinfo{pages}{5897} (\bibinfo{year}{1985}).

\bibitem[{\citenamefont{Zahn et~al.}(2002)\citenamefont{Zahn, Schilling, and
  Kast}}]{zahn_2002}
\bibinfo{author}{\bibfnamefont{D.}~\bibnamefont{Zahn}},
  \bibinfo{author}{\bibfnamefont{B.}~\bibnamefont{Schilling}},
  \bibnamefont{and} \bibinfo{author}{\bibfnamefont{S.~M.} \bibnamefont{Kast}},
  \bibinfo{journal}{J. Phys. Chem. B} \textbf{\bibinfo{volume}{106}},
  \bibinfo{pages}{10725} (\bibinfo{year}{2002}).

\bibitem[{\citenamefont{Wolf et~al.}(1999)\citenamefont{Wolf, Keblinski,
  Phillpot, and Eggebrecht}}]{wolf_1999}
\bibinfo{author}{\bibfnamefont{D.}~\bibnamefont{Wolf}},
  \bibinfo{author}{\bibfnamefont{P.}~\bibnamefont{Keblinski}},
  \bibinfo{author}{\bibfnamefont{S.~R.} \bibnamefont{Phillpot}},
  \bibnamefont{and}
  \bibinfo{author}{\bibfnamefont{J.}~\bibnamefont{Eggebrecht}},
  \bibinfo{journal}{J. Chem. Phys.} \textbf{\bibinfo{volume}{110}},
  \bibinfo{pages}{8254} (\bibinfo{year}{1999}).

\bibitem[{\citenamefont{Ewald}(1921)}]{ewald_1921}
\bibinfo{author}{\bibfnamefont{P.}~\bibnamefont{Ewald}}, \bibinfo{journal}{Ann.
  Phys.} \textbf{\bibinfo{volume}{369}}, \bibinfo{pages}{253}
  (\bibinfo{year}{1921}).

\bibitem[{\citenamefont{Demontis et~al.}(2001)\citenamefont{Demontis, Spanu,
  and Suffritti}}]{demontis_2001}
\bibinfo{author}{\bibfnamefont{P.}~\bibnamefont{Demontis}},
  \bibinfo{author}{\bibfnamefont{S.}~\bibnamefont{Spanu}}, \bibnamefont{and}
  \bibinfo{author}{\bibfnamefont{G.~B.} \bibnamefont{Suffritti}},
  \bibinfo{journal}{J. Chem. Phys.} \textbf{\bibinfo{volume}{114}},
  \bibinfo{pages}{7980} (\bibinfo{year}{2001}).

\bibitem[{\citenamefont{Fennell and Gezelter}(2006)}]{fennell_2006}
\bibinfo{author}{\bibfnamefont{C.~J.} \bibnamefont{Fennell}} \bibnamefont{and}
  \bibinfo{author}{\bibfnamefont{J.~D.} \bibnamefont{Gezelter}},
  \bibinfo{journal}{J. Chem. Phys.} \textbf{\bibinfo{volume}{124}},
  \bibinfo{pages}{234104} (\bibinfo{year}{2006}).

\bibitem[{\citenamefont{Denesyuk and Weeks}(2008)}]{denesyuk_2008}
\bibinfo{author}{\bibfnamefont{N.~A.} \bibnamefont{Denesyuk}} \bibnamefont{and}
  \bibinfo{author}{\bibfnamefont{J.~D.} \bibnamefont{Weeks}},
  \bibinfo{journal}{J. Chem. Phys} \textbf{\bibinfo{volume}{128}},
  \bibinfo{pages}{124109} (\bibinfo{year}{2008}).

\bibitem[{\citenamefont{Levitt et~al.}(1995)\citenamefont{Levitt, Hirshberg,
  Sharon, and Daggett}}]{levitt_1995}
\bibinfo{author}{\bibfnamefont{M.}~\bibnamefont{Levitt}},
  \bibinfo{author}{\bibfnamefont{M.}~\bibnamefont{Hirshberg}},
  \bibinfo{author}{\bibfnamefont{R.}~\bibnamefont{Sharon}}, \bibnamefont{and}
  \bibinfo{author}{\bibfnamefont{V.}~\bibnamefont{Daggett}},
  \bibinfo{journal}{Comp. Phys. Comm.} \textbf{\bibinfo{volume}{91}},
  \bibinfo{pages}{215} (\bibinfo{year}{1995}).

\bibitem[{\citenamefont{Beck et~al.}(2005)\citenamefont{Beck, Armen, and
  Daggett}}]{beck_2005}
\bibinfo{author}{\bibfnamefont{D.~A.~C.} \bibnamefont{Beck}},
  \bibinfo{author}{\bibfnamefont{R.~S.} \bibnamefont{Armen}}, \bibnamefont{and}
  \bibinfo{author}{\bibfnamefont{V.}~\bibnamefont{Daggett}},
  \bibinfo{journal}{Biochem.} \textbf{\bibinfo{volume}{44}},
  \bibinfo{pages}{609} (\bibinfo{year}{2005}).

\bibitem[{\citenamefont{Hansen and McDonald}(1975)}]{hansen_1975}
\bibinfo{author}{\bibfnamefont{J.~P.} \bibnamefont{Hansen}} \bibnamefont{and}
  \bibinfo{author}{\bibfnamefont{I.~R.} \bibnamefont{McDonald}},
  \bibinfo{journal}{Phys. Rev. A} \textbf{\bibinfo{volume}{11}},
  \bibinfo{pages}{2111} (\bibinfo{year}{1975}).

\bibitem[{\citenamefont{Nos\'{e}}(1984)}]{nose_1984}
\bibinfo{author}{\bibfnamefont{S.}~\bibnamefont{Nos\'{e}}},
  \bibinfo{journal}{Mol. Phys.} \textbf{\bibinfo{volume}{52}},
  \bibinfo{pages}{255} (\bibinfo{year}{1984}).

\bibitem[{\citenamefont{Hoover}(1985)}]{hoover_1985}
\bibinfo{author}{\bibfnamefont{W.~G.} \bibnamefont{Hoover}},
  \bibinfo{journal}{Phys. Rev. A} \textbf{\bibinfo{volume}{31}},
  \bibinfo{pages}{1695} (\bibinfo{year}{1985}).

\bibitem[{\citenamefont{Wu et~al.}(2006)\citenamefont{Wu, Tepper, and
  Voth}}]{wu_2006}
\bibinfo{author}{\bibfnamefont{Y.}~\bibnamefont{Wu}},
  \bibinfo{author}{\bibfnamefont{H.~L.} \bibnamefont{Tepper}},
  \bibnamefont{and} \bibinfo{author}{\bibfnamefont{G.~A.} \bibnamefont{Voth}},
  \bibinfo{journal}{J. Chem. Phys.} \textbf{\bibinfo{volume}{124}},
  \bibinfo{pages}{024503} (\bibinfo{year}{2006}).

\bibitem[{\citenamefont{Raabe and Sadus}(2007)}]{raabe_2007}
\bibinfo{author}{\bibfnamefont{G.}~\bibnamefont{Raabe}} \bibnamefont{and}
  \bibinfo{author}{\bibfnamefont{R.~J.} \bibnamefont{Sadus}},
  \bibinfo{journal}{J. Chem. Phys.} \textbf{\bibinfo{volume}{126}},
  \bibinfo{pages}{044701} (\bibinfo{year}{2007}).

\bibitem[{\citenamefont{Darden et~al.}(1993)\citenamefont{Darden, York, and
  L.Pedersen}}]{darden_1993}
\bibinfo{author}{\bibfnamefont{T.}~\bibnamefont{Darden}},
  \bibinfo{author}{\bibfnamefont{D.}~\bibnamefont{York}}, \bibnamefont{and}
  \bibinfo{author}{\bibnamefont{L.Pedersen}}, \bibinfo{journal}{J. Chem. Phys.}
  \textbf{\bibinfo{volume}{98}}, \bibinfo{pages}{10089} (\bibinfo{year}{1993}).

\bibitem[{\citenamefont{Karlstr\"{o}m et~al.}(2008)\citenamefont{Karlstr\"{o}m,
  Stenhammer, and Linse}}]{karlstrom_2008}
\bibinfo{author}{\bibfnamefont{G.}~\bibnamefont{Karlstr\"{o}m}},
  \bibinfo{author}{\bibfnamefont{J.}~\bibnamefont{Stenhammer}},
  \bibnamefont{and} \bibinfo{author}{\bibfnamefont{P.}~\bibnamefont{Linse}},
  \bibinfo{journal}{J. Phys.: Condens. Matter} \textbf{\bibinfo{volume}{20}},
  \bibinfo{pages}{494204} (\bibinfo{year}{2008}).

\bibitem[{\citenamefont{Takahashi et~al.}(2011)\citenamefont{Takahashi, Narumi,
  and Yasouko}}]{takahashi_2011}
\bibinfo{author}{\bibfnamefont{K.}~\bibnamefont{Takahashi}},
  \bibinfo{author}{\bibfnamefont{T.}~\bibnamefont{Narumi}}, \bibnamefont{and}
  \bibinfo{author}{\bibfnamefont{K.}~\bibnamefont{Yasouko}},
  \bibinfo{journal}{J. Chem. Phys.} \textbf{\bibinfo{volume}{134}},
  \bibinfo{pages}{174112} (\bibinfo{year}{2011}).

\bibitem[{\citenamefont{Rodgers and Weeks}(2008)}]{rodgers_2008}
\bibinfo{author}{\bibfnamefont{J.~M.} \bibnamefont{Rodgers}} \bibnamefont{and}
  \bibinfo{author}{\bibfnamefont{J.~D.} \bibnamefont{Weeks}},
  \bibinfo{journal}{PNAS} \textbf{\bibinfo{volume}{105}},
  \bibinfo{pages}{19136} (\bibinfo{year}{2008}).

\bibitem[{\citenamefont{Wu and Brooks}(2005)}]{wu_2005}
\bibinfo{author}{\bibfnamefont{X.}~\bibnamefont{Wu}} \bibnamefont{and}
  \bibinfo{author}{\bibfnamefont{B.~R.} \bibnamefont{Brooks}},
  \bibinfo{journal}{J. Chem. Phys.} \textbf{\bibinfo{volume}{122}},
  \bibinfo{pages}{044107} (\bibinfo{year}{2005}).

\end{thebibliography}

\end{document}